\documentclass[useAMS,usenatbib]{mn2e} \usepackage{latexsym}
\usepackage{times} \usepackage{subeqn} \usepackage{graphics}
\usepackage[german,english,american]{babel} \newcommand{\apj}{ApJ}
\newcommand{\apjl}{ApJL} \newcommand{\nat}{Nature}
\newcommand{\aap}{A\&A} \newcommand{\mnras}{MNRAS}
\newcommand{\aaps}{A\&AS}

\def\lesssim{\mathrel{\hbox{\rlap{\hbox{\lower4pt\hbox{$\sim$}}}\hbox{$<$}}}}
\def\gtrsim{\mathrel{\hbox{\rlap{\hbox{\lower4pt\hbox{$\sim$}}}\hbox{$>$}}}}

\title[Constraints on jet X-rays]{Constraints on the role of
  synchrotron X-rays from jets of accreting black holes}

\author[S.~Heinz]{S.~Heinz\\Center for Space Research, Massachusetts
Institute of Technology, 77 Massachusetts Avenue, Cambridge, MA
02139\\ Chandra Fellow}

\begin{document}
\date{1 September 2004} \maketitle

\begin{abstract}
We present an extension of the scale invariance formalism by
\citet{heinz:03a} which includes the effects of radiative cooling on
the X-ray synchrotron emission from core dominated jets and outflows.
We derive the scaling relations between the radio luminosity $L_{\rm
r}$, the synchrotron X-ray luminosity $L_{\rm x}$, the mass $M$, and
the accretion rate $\dot{m}$ of an accreting black hole.  We argue
that the inclusion of synchrotron losses into the scaling relations is
required by the data used in \cite{merloni:03} to define the
``fundamental plane'' correlation between $L_{\rm r}$, $L_{\rm x}$,
and $M$ in AGNs amd Galactic X-ray binaries. We then fit the new
scaling relations to the ``fundamental plane'' correlation.  This
allows us to derive statistical constraints on the contribution from
jet synchrotron emission to the X-rays from black holes, or,
alternatively, on the particle acceleration scheme at work in these
jets.  We find that, in order for jet synchrotron X-rays to be
consistent with the observed $L_{\rm r} - L_{\rm x} - M$ correlation,
the global particle spectrum must be both {\em steep}: $f(\gamma)
\propto \gamma^{-(3.4^{+0.6}_{-0.5})}$ and {\em unbroken} down to
energies well below the radio synchrotron regime.  Such a steep
particle distribution would imply steeper X-ray spectra than observed
in most sources contributing to the correlation with avalailable X-ray
spectral index measurements.  We suggest that, unless some assumptions
of the scale invariance hypothesis are broken, another emission
mechanism (likely either disk radiation and/or inverse Compton
scattering in the jet) contributes significantly to the X-rays of a
sizeable fraction of the sources.
\end{abstract}

\section{Introduction}
For many years, the standard paradigm for the X-ray emission in
accreting black hole systems has been that the X-rays are produced in
the central accretion flow
\citep[e.g.,][]{shakura:73,shapiro:76,narayan:94}, while radio
emission was traditionally attributed to a radio jet. Only in the most
powerful, beamed jet sources was X-ray synchrotron and inverse Compton
emission from the jet invoked to explain the featureless powerlaw
spectra of Blazars and BL-Lacs \citep[e.g.][]{fossati:98}.

In recent years, however, this paradigm has been called into
question. Beginning with a re--evaluation of the emission processes in
Galactic X-ray binaries (XRBs), an alternative view has been proposed
to interpret a large part of the spectrum (including X-rays) of black
holes at moderate to low accretion rates entirely as due to the
emission from the jet \citep{markoff:01,markoff:03,falcke:04}.

It has been difficult to find unique signatures that would allow us to
distinguish between these two alternatives. Broad band spectral
modeling is the most promising approach, but so far it has not been
shown conclusively that either of the two approaches fails.  The
strong correlation between the radio and the X-ray emission in X-ray
binaries that has been found recently \citep{corbel:03,gallo:03} can
be interpreted in terms of both pictures. Similarly, the correlation
between radio luminosity $L_{\rm r}$, X-ray luminosity $L_{\rm x}$,
and black hole mass $M$ is consistent with both pictures
\citep{merloni:03,falcke:04}.

In order to investigate the relation between $L_{\rm xr}$, $L_{\rm
x}$, and $M$, \citet{heinz:03a} derived robust and very general
scaling relations between these observables which are {\em
independent} of the underlying jet model
\citep[e.g.][]{blandford:78,falcke:95}, provided that jet physics is
scale invariant under changes in black hole mass $M$ and accretion
rate $\dot{m}=\dot{M}/\dot{M}_{\rm Edd}$ (where $\dot{M}_{\rm Edd}$ is
the Eddington accretion rate).  The striking agreement with the
subsequently found ``fundamental plane'' relation of accreting black
holes \citep{merloni:03,falcke:04} seems to confirm the scale
invariance hypothesis\footnote{While the influence of interactions
between the jet and the environment must affect the radiative
signature of jets on larger scales \citep{heinz:02}, we note that the
scaling relation predicted by \cite{heinz:03a} for optically thin
synchrotron emission agrees very well with the tentative mass and
accretion rate dependence of the optically thin radio luminosity of
extended jets found by \cite{sams:96}.}.

The original treatment of \citet{heinz:03a} did, however, leave out
some critical physical ingredients required to describe the X-ray
synchrotron emission from jets, namely, radiative cooling of the
particle distribution, which alters the spectral shape at higher
energies. In this paper, we will present an extension of the formalism
developed in \citet{heinz:03a} that allows us to include radiative
cooling (\S\ref{sec:cooling} and \S\ref{sec:appendix}). We then
proceed to investigate the data of the ``fundamental plane'' relation
for signs of radiative cooling and discuss limits we can put on the
contribution of X-ray synchrotron radiation from jets to the spectra
of AGNs and XRBs by fitting the derived scaling relations to the
fundamental plane correlation (\S\ref{sec:comparison}). Section
\ref{sec:conclusions} summarizes the conclusions we draw from this
analysis.

\section{Scale invariance in jets with radiative losses} 
\label{sec:cooling}
In \citet{heinz:03a} we derived the scaling relations between
synchrotron radio luminosity $L_{\rm r}$, black hole mass $M$, and
accretion rate $\dot{M}$.  Because radiative losses introduce another
scale into the problem, we chose to ignore the effect of synchrotron
cooling on the particle distribution, given the scope of the paper.
As we will show below, it is possible to incorporate these effects
without loss of generality, because the scale introduced by cooling
depends itself only on the mass and accretion rate of the black hole
and does not enter into the dynamical properties of the jet plasma at
all.  This is because cooling affects primarily the highest energy
particles, while most of the jet kinetic energy and thrust are carried
by the low energy particles (in other words, jets are believed to be
radiatively inefficient).

\subsection{The scale invariance hypothesis for cooled jets}
In order to include the effects of radiative cooling into the scale
invariance formalism \citep{heinz:03a}, we shall briefly review the
aspects of the scale invariance ansatz relevant to this article. For a
detailed discussion, we refer the interested reader to the original
article.

The scale invariance ansatz assumes that there is only one relevant
scale in the problem of jet dynamics (at least in the cores of jets
which we are focusing on here): the gravitational radius $r_{\rm g} =
GM/c^2$.  It also assumes that the structure of all core jets$^{1}$
from black holes is the same when set in relation to $r_{\rm g}$.
Mathematically, the scale invariance assumption can be formulated as
the requirement that any dynamical quantity $f$ relevant to
calculating the synchrotron emission from jets (such as the
normalization of the powerlaw distribution of electrons, the magnetic
field, or the particle pressure) can be decomposed into two functions
$\phi_f(M,\dot{m})$ and $\psi_f(\chi)$, such that
\begin{equation}
  f(M,\dot{m},r) = \phi_f(M,\dot{m})\psi_f(\chi)
\label{eq:scaleinvariant}
\end{equation}
where the definition $\chi = r/r_{\rm g}$ implies that $\psi_f$
depends only on scale--free coordinates $(x,y,z)/r_{\rm g}$. 

The functions $\phi_f$ define the normalization of $f$ at the base of
the jet (as provided by the accretion disk), while the structure
functions $\psi_f$ describe the variation of $f$ along the jet,
relative to its value at the base of the jet.  Any non-trivial
dependence of $f$ on the black hole spin $a$ is assumed to be implicit
in the functions $\psi_{f}$ and will not be discussed further.  For
consistency, we will also assume that the jet velocity $\beta\Gamma$
is independent of black hole mass and accretion rate, such that
$\beta\Gamma = \psi_{\beta\Gamma}(\chi)$.  It is still not known what
the true jet plasma velocities are, especially close to the base of
the jet, but future constraints will hopefully shed some light on this
subject.  If this assumption is not satisfied, the formalism can
easily be altered to include such an $M$ or $\dot{m}$ dependence as
long as the four-velocity $\beta\Gamma$ can be written in the form of
eq.~(\ref{eq:scaleinvariant}).

Given these assumptions, we can proceed to include radiative cooling
into the treatment.  We will assume that a fraction of the jet
particles are accelerated at some arbitrary time $t_0$ (measured in
the jet frame) into a powerlaw such that
\begin{equation}
  f(\gamma,t_0) = \phi_{C} \gamma^{-p}
  \label{eq:powerlaw}
\end{equation}
and then evolve through adiabatic and radiative cooling as they move
downstream from the acceleration region.  

While it is not necessary to assume that this acceleration takes place
in one particular location (it can be spatially extended, in which
case the integral in eq.~(\ref{eq:fnu}) can be performed in a
piece-wise fashion for every acceleration zone), it will be necessary
to make the assumption that the spatial distribution of the
acceleration (i.e., the various positions of the acceleration regions
measured in units of gravitational radii $r_g$) and the shape of the
powerlaw are invariant under changes of $M$ and $\dot{m}$ (in line
with the assumptions necessary to derive the original scale invariance
properties in \citealt{heinz:03a}).  This implies that the injection
time in the jet frame $t_0$ is proportional to $t_0 \propto r_{\rm g}
\propto M$.  This is reasonable if the particle acceleration is
associated with a hydrodynamic process, such as a recollimation shock,
since the underlying hydrodynamics is also assumed to be scale
invariant.

Furthermore, we will assume that the particle pressure is dominated by
the low energy end of the particle distribution, which implies that
the slope of the particle distribution produced by the acceleration
process is steeper than $p \gtrsim 2$.  Classical Fermi I shock
acceleration in strong, non-relativistic shocks produces spectra with
index $p \gtrsim 2$ \citep{blandford:78,bell:78,drury:82}, while the
spectral index produced in strong relativistic shocks is somewhat
steeper, $p \sim 2.3$ \citep{gallant:99b,achterberg:01}.  

These indices agree nicely with the observed optically thin radio
synchrotron spectra of extended jets with spectral indices of order
$0.5 \lesssim \alpha_{\rm r} \lesssim 0.7$ (where the radio spectral
index is defined as $\alpha_{\rm r} \equiv -\partial{(\log{F_{\rm
r}})}/\partial{(\log{\nu})}$).  For this reason, we will assume in the
following that the fiducial spectral index of the uncooled electron
distribution falls into the range $2 \leq p \leq 2.3$.  The fiducial
uncooled synchrotron spectral index falls into the range $0.5 \leq
\alpha \leq 0.65$.  More recent efforts studying particle acceleration
in jets are focusing on continuous particle acceleration mechanisms
like reconnection \citep[e.g.][]{larrabee:03}, which might be
necessary to explain the lack of synchrotron cooling signatures in
some extended AGN jets like 3C273 \citep{jester:01}.

\subsection{Scaling relations}
With the basic set of assumptions in place, we proceed to derive the
scaling relations for the X-ray synchrotron emission from jet cores
{\em including} the effects of radiative cooling on the particle
distribution.  The complete mathematical derivation used in the
remainder of this paper is presented in the appendix.  For the casual
reader, we will present a sketch of the derivation in the following
paragraphs.

\begin{figure}
\resizebox{\columnwidth}{!}{\includegraphics{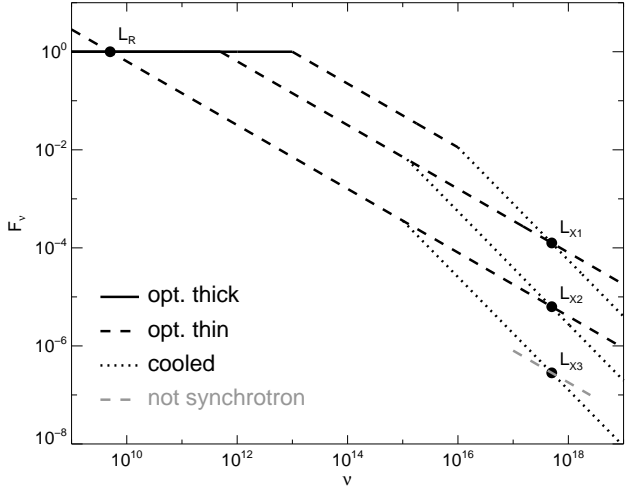}}
\caption{Sketch of viable synchrotron spectra in different regimes for
  three different measured radio-to-X-ray ratios. Optically thick
  parts are shown as solid lines, optically thin, uncooled parts are
  shown as dashed lines, optically thin, radiatively cooled parts are
  shown as dotted lines.  The grey dashed line through L$_{\rm X3}$
  shows an X-ray spectrum that is flatter than the global spectrum,
  which would require a concave spectrum and would be inconsistent
  with simple synchrotron X-rays.\label{fig:sketch}}
\end{figure}

Synchrotron cooling of a population of electrons in the presence of a
source of fresh powerlaw particles (which is the appropriate model for
a continuous jet) will produce a cooling break in the global spectrum,
as sketched in Fig.~\ref{fig:sketch}.  Given the field strength $B$,
the characteristic ``cooling energy'' of this break in the electron
spectrum is \citep[e.g.][also, see eqs.~(\ref{eq:characteristic}) and
(\ref{eq:gamma}) for a derivation of this expression]{rybicki:79}
$\gamma_{\rm b} \sim \left[{A_{0}B^{2}t_{\rm jet}}\right]^{-1}$, where
$t_{\rm jet}\sim r_{\rm jet}/v_{\rm jet} \propto M$ is the
characteristic travel time of the electrons.  Using
eq.~(\ref{eq:scaleinvariant}), this expression becomes $\gamma_{\rm b}
\propto {\phi_{\rm B}(M,\dot{m})}^{-2}M^{-1}$.  The characteristic
synchrotron frequency corresponding to this electron energy is
\begin{equation}
  \nu_{\rm b} \propto B\gamma_{\rm b}^2 \propto
\left[\phi_{\rm B}(M,\dot{m})\right]^{-3}M^{-2}
\label{eq:nub}
\end{equation}  

We will assume that the spectrum has a break at $\nu_{\rm b}$ with
powerlaw index $\alpha_{\rm x}$ above $\nu_{\rm b}$ and with the
standard powerlaw index $\alpha=(p-1)/2$ below $\nu_{\rm b}$ (see
sketch in Fig.~\ref{fig:sketch}).  Given the expression for $\nu_{\rm
b}$ from eq.~(\ref{eq:nub}), the flux at a fixed frequency $\nu_{\rm
x}$ above the cooling break $\nu_{\rm b}$ is a simple function of the
flux $F_{\nu_{0}}$ at a fixed frequency $\nu_0$ below the cooling
break, the black hole mass $M$, and the accretion rate $\dot{m}$:
\begin{equation}
  F(\nu_{\rm x}) = F_{\nu_0}\left(\frac{\nu_{\rm 0}}{\nu_{\rm
    b}}\right)^{(p-1)/2} \left(\frac{\nu_{\rm b}}{\nu_{\rm
    x}}\right)^{-\alpha_{\rm x}}
\end{equation}
The {\em uncooled}, optically thin jet flux $F(\nu_0)$ depends on $M$
and $\dot{m}$ as follows \citep{heinz:03a}:
\begin{equation}
  F(\nu_0) \propto
  M^{3}\phi_{B}^{\frac{p+1}{2}}(M,\dot{m})\phi_{C}(M,\dot{m})
  \label{eq:uncooled}
\end{equation}
With the expression for $F(\nu_{\rm x})$ as a function of $F(\nu_0)$,
$M$, and $\dot{m}$ from above, we have
\begin{equation}
  F(\nu_{\rm x}) \propto M^{2+p-\alpha_{\rm
  x}}\phi_{B}^{2p-1-3\alpha_{\rm x}}\phi_{C}
\end{equation}
Writing this expression in differential form, we arrive at the scaling
relation derived in full generality in the appendix in
eq.~(\ref{eq:axim})
and eq.~(\ref{eq:aximdot})

We expect the functions $\phi_{B}(M,\dot{m})$ and
$\phi_{C}(M,\dot{m})$ to be of powerlaw type, given typical accretion
disk scenarios; \citet{heinz:03a} and \citet{merloni:03} discuss the
possible functional forms of $\phi_{B}$ and $\phi_{C}$.  In this case,
the logarithmic derivatives $\xi_{\rm xM}$ in eq.~(\ref{eq:axim}) and
$\xi_{\rm x\dot{m}}$ in eq.~(\ref{eq:aximdot}) are constant and the
scaling relation between $F_{\rm x}$ and $M$ and between $F_{\rm x}$
and $\dot{m}$ assumes powerlaw form:
\begin{eqnarray}
  F_{\rm x} & \propto &
  M^{\xi_{xM}}\dot{m}^{\xi_{x\dot{m}}} \\ \xi_{\rm
  xM} & = & 2 - 2\alpha_{\rm x} + p +\frac{\partial
  \log{(\phi_{C}})}{\partial \log{M}} \nonumber \\ & & + (2p - 1 -
  3\alpha_{\rm x})\frac{\partial \log{(\phi_{B})}}{\partial \log{M}}
  \label{eq:xim} \\
  \xi_{\rm x\dot{m}} & = & \frac{\partial \log{(\phi_{C}})}{\partial
  \log{\dot{m}}} + (2p - 1 - 3\alpha_{\rm x})\frac{\partial
  \log{(\phi_{B})}}{\partial \log{\dot{m}}}
  \label{eq:ximdot}
\end{eqnarray}

Expressing the scaling relation this way is useful because $p$ and
$\alpha_{\rm x}$ are actually {\em observables}.  In other words, we
can substitute observable quantities (in this case the X-ray and the
particle spectral indices $\alpha_{\rm x}$ and $p$ respectively) for
the unknown details of particle acceleration and
cooling\footnote{\citet{heinz:03a} used the same approach to
substitute $p$ and the observable radio spectral index $\alpha_{\rm
r}$ for the unknown details of the jet structure.}.  It is important
to keep in mind that $p$ is the electron spectral index produced by
the acceleration process, i.e. {\em below} the cooling break and thus
{\em without} any modification due to cooling. $p$ is therefore not
observable via the X-ray spectral index. Rather, optically thin, {\em
uncooled} radio, sub-millimeter, or IR emission should be used to
infer $p$, since this radiation should not be affected by synchrotron
losses.

In the following we will use the same canonical functional forms for
$\phi_{B}$ and $\phi_{C}$ used in \citep{heinz:03a}.  Namely, we
assume that the underlying accretion flow is mechanically
cooled\footnote{$\dot{m}$ is the accretion rate at the inner edge of
the disk, since this is where the jets are formed and is thus well
defined even for ADIOS-- \citep{blandford:99} and CDAF--
\citep{narayan:00,quataert:00} type flows.}, in which case the
accretion flow itself is scale invariant.  Then
\begin{equation}
   \phi_C = {\phi_B}^2 = \dot{m}M^{-1}
\label{eq:phi_c}
\end{equation}
This prescription is also valid for coronae above gas
pressure-dominated, radiatively efficient \citep{shakura:73} accretion
disks \citep[the standard scenario for the origin of powerlaw X-ray
emission and the launching point for jets from geometrically thin
accretion disks, e.g.][]{merloni:02}.  We then have
\begin{eqnarray}
  \xi_{{\rm x}M} & = & \frac{3 - \alpha_{\rm x}}{2} \label{eq:xim2} \\
  \xi_{{\rm x}\dot{m}} & = & p + \frac{1 - 3\alpha_{\rm x}}{2}
  \label{eq:ximdot2}
\end{eqnarray}
These are the expressions we will use in the remainder of this paper
to fit the ``fundamental plane'' correlation.  In the uncooled limit
of $\alpha_{\rm x} = (p-1)/2$, they reduce to the expressions derived
for the optically thin case from eq.~(\ref{eq:uncooled}). In the
canonical case of $p=2$, we have $F_{\rm x} \propto M^{(3-\alpha_{\rm
x})/2}{\dot{m}}^{(5-3\alpha_{\rm x})/2}$.

Since the ``fundamental plane'' correlation is measured between the
radio and X-ray luminosities and the black hole mass, we should
convert the scaling relations to those coordinates.  From
\cite{heinz:03a}, the radio flux follows the scaling relation
\begin{equation}
    F_{\rm r} \propto M^{\frac{2p + 13 - (2+p)\alpha_{\rm r}}{8+2p}}
      {\dot{m}}^{\frac{2p + 13 + (p+6)\alpha_{\rm r}}{2\left(p+4\right)}}
\end{equation}
Combining this expression with eqs.~(\ref{eq:xim2}) and
(\ref{eq:ximdot2}), we finally arrive at the desired relation between
$F_{\rm r}$, $M$, and the synchrotron X-ray emission $F_{\rm x}$:
\begin{eqnarray}
  F_{\rm r} & \propto & M^{\xi_{\rm RM}}{F_{\rm x}}^{\xi_{\rm RX}}\\
  \xi_{\rm RM} & = & \frac{\left(2p + 13 - \left(2 +
  p\right)\alpha_{\rm r}\right)\left(p - 1 - \alpha_{\rm x}\right) -
  \left(2\alpha_{\rm
  r}\right)}{\left(p+4\right)\left(2p+1-3\alpha_{\rm x}\right)}
  \label{eq:xirm} \\
  \xi_{\rm RX} & = & \frac{2p + 13 + \left(p + 6\right)\alpha_{\rm
  r}}{\left(p+4\right)\left(2p + 1 - 3\alpha_{\rm
  x}\right)}\label{eq:xirx}
\end{eqnarray}
This can now be compared directly to the measured correlation coefficients
$\xi_{\rm RM,FP}=0.78$ and $\xi_{\rm RX,FP}=0.6$.

\section{Constraints on the X-ray emission mechanism}
\label{sec:comparison}
The ``fundamental plane'' correlation has the form
\begin{equation}
  \log {L_{\rm r}} = 0.78 \log{M} + 0.6 \log{L_{\rm x}} + 7.33
  \label{eq:fundamentalplane}
\end{equation}
\cite{merloni:03} interpreted this correlation in terms of the scale
invariance model.  From the observed correlation, it is unclear
whether the source of the X-rays is, in fact, synchrotron emission
from the jet, or whether it is emission from an inefficiently
accreting disk; both can produce similar correlation coefficients in
the context of the scale invariance model.  The scaling relations for
inefficient disk X-rays do produce a better fit to the fundamental
plane, but \cite{merloni:03} noted that the effects of synchrotron
losses on the synchrotron X-ray emission from the jet had not been
included in their treatment and might improve the fit to the
fundamental plane data.  Given the analysis presented in the previous
section, we can now proceed to include these effects and test whether
they improve the fit or not.

\subsection{Spectral evidence for synchrotron cooling}
\label{sec:cooling2}
Before applying the scaling relations derived in eqs.~(\ref{eq:xirm})
and (\ref{eq:xirx}) to the ``fundamental plane'' sample, we shall
present a brief argument that the inclusion of cooling into the scale
invariance relations presented above is not just useful, but actually
necessary to properly interpret the fundamental plane data in the
context of synchrotron jet X-rays.

First, we will check the global radio-X-ray spectra of individual
sources against the hypothesis that the X-rays originate in the jet as
optically thin, {\em uncooled} synchrotron emission.  In this case,
the X-ray spectrum should continue from the X-rays to lower
frequencies at the standard synchrotron spectral index $\alpha_{\rm x}
= (p-1)/2$, until it reaches the turnover frequency where the jet
becomes optically thick to synchrotron self absorption (see
Fig.~\ref{fig:sketch}).  At this point, the spectrum becomes
flat. This implies that the spectrum is convex at every point between
the radio and X-ray bands.

\begin{figure}
\resizebox{\columnwidth}{!}{\includegraphics{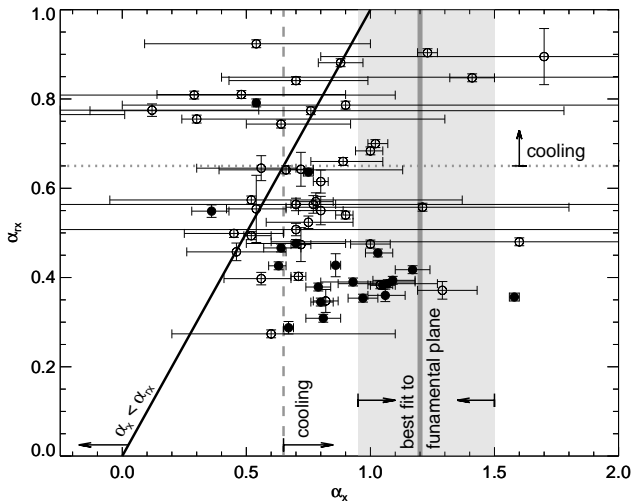}}
\caption{Plot of the measured radio-X-ray spectral index $\alpha_{\rm
  rx}$ against the measured X-ray spectral index $\alpha_{\rm x}$ for
  the AGNs in the ``fundamental plane'' sample where this information
  is available.  {\em left of black line}: X-ray spectrum flatter than
  radio-to-X-ray spectrum, $\alpha_{\rm x} < \alpha_{\rm rx}$,
  difficult to reconcile with global synchrotron spectrum; {\em light
  grey area}: X-ray spectral indices from best fit to ``fundamental
  plane'' correlation (see Fig.~\ref{fig:alpha_p}); {\em dashed and
  dotted grey lines}: cooling or steep injection spectrum necessary to
  the right of and above these lines; {\em filled dots}: sources with
  X-ray luminosity $L_{\rm x} > 10^{-3}\,L_{\rm Edd}$, {\em empty
  dots}: sources with $L_{\rm x} < 10^{-3}\,L_{\rm
  Edd}$.\label{fig:spectra}}
\end{figure}

Therefore, if the global spectral index $\alpha_{\rm rx}$ is steeper
than the fiducial uncooled X-ray spectral index $\alpha_{\rm
X}=(p-1)/2$, radiative cooling {\em must} have steepened the jet
synchrotron spectrum at X-ray energies (this corresponds to point $\rm
L_{\rm X3}$ in Fig.~\ref{fig:sketch}).  Only a cooling break/cutoff
can reduce the X-ray flux to the observed level (even if the radio
spectrum is optically thick)\footnote{If the X-rays originate in the
disk, this limit becomes even stronger, since the jet-spectral index
is then even steeper, thus also implying the action of radiative
cooling.}.  This constraint corresponds to the horizontal dashed grey
line in Fig.~\ref{fig:spectra}, and a number of sources lie above this
line, showing that radiative cooling must be included in the treatment
of these sources.  Note that relativistic beaming and uncorrelated
radio/X-ray variability do not change this conclusion.

On the other hand, if the global spectral index $\alpha_{\rm rx,FP}$
is flatter than the fiducial uncooled synchrotron spectrum (point $\rm
L_{X1}$ in Fig.~\ref{fig:sketch}, all points below the dashed grey
line in Fig.~\ref{fig:spectra}), we cannot say anything about the
action of radiative cooling. All we can conclude in this case is that
either the radio synchrotron emission must be self-absorbed (thus
flattening the global synchrotron spectrum), or the X-rays must be
from the disk (thus increasing the X-ray flux relative to the radio).

For a sub-set of the sources in the ``fundamental plane'' sample, the
X-ray spectral indices $\alpha_{\rm x}$ have actually been measured,
though for some sources with very large error bars.  All of these
sources are AGNs, as the RXTE-ASM data used in compiling the
``fundamental plane'' sample are not sufficient to determine the
spectral slope between 2 and 10 keV.  However, we note that for XRBs
in the low-hard state, where the radio-X-ray correlation is observed
to hold, the X-ray powerlaw spectra are typically hard - with $0.5
\lesssim \alpha_{\rm x} \lesssim 1.0$.  Most of what will be discussed
regarding the AGN with measured $\alpha_{\rm x}$ will also hold for
these sources.

We have plotted the global and X-ray spectral indices of all AGNs in
the sample where $\alpha_{\rm x}$ has been measured in
Fig.~\ref{fig:spectra}.  This figure shows several interesting things:
\begin{itemize}
\item{Some sources have global spectra steeper than $\alpha_{\rm rx} >
  0.65$ (above the dotted grey line), which implies that they are
  affected by cooling or have unusually steep particle spectra.
  Because XRBs are generally less radio loud than AGNs, all XRBs fall
  well below this line (they are not shown on this plot for lack of
  $\alpha_{\rm x}$ measurements), thus, this observational constraint
  on the importance of cooling is not available for XRBs.}
\item{Most of the sources in the plot have X-ray spectra {\em steeper}
  than the fiducial uncooled synchrotron spectrum of $\alpha_{\rm X} >
  0.65$ (right of dashed grey line). This indicates that, if the
  X-rays are due to synchrotron emission, (1) radiative cooling should
  have altered their spectrum {\em or} (2) the particle injection
  spectrum must be steeper than typical Fermi I shock acceleration.}
\item{About 25\% of the sources in the plot actually fall into the
  area left of the solid black line where the X-ray spectral index is
  {\em flatter} than the global spectral index (corresponding to the
  dashed grey line through point $\rm L_{X3}$ in
  Fig.~\ref{fig:sketch}), i.e., $\alpha_{\rm x} < \alpha_{\rm rx}$.
  These sources have concave global spectra and the X-rays from these
  sources {\em cannot} be due to X-ray synchrotron emission from the
  jet, unless a particle acceleration scheme is at work that produces
  concave electron spectra.  Again, all XRBs lie well below this line,
  so this observational constraint on the nature of the emission
  mechanism is not available for XRBs.  We note that the error bars on
  $\alpha_{\rm x}$ for all of the sources left of the line are large.
  Thus, it is prudent not to draw any firm conclusions from this
  seeming discrepancy at this point and await better observational
  determinations of $\alpha_{\rm x}$ for these sources.}
\end{itemize}

In summary, we find that radiative cooling must be at work in most of
the AGNs contributing to the ``fundamental plane'' sample, unless the
particle spectra are intrinsically somewhat steeper than those
produced in Fermi I acceleration and typically observed in optically
thin radio synchrotron spectra of relativistic jets.  This underlines
the need for radiative cooling to be included in properly modeling the
``fundamental plane'' correlation.

\subsection{Fitting the scaling relations to the ``fundamental
  plane'' data set} 

We will now compare the new scaling relations to the fundamental plane
correlation.  We will assume in the following that the radio emission
is unaffected by radiative cooling.  Therefore, if the X-rays
originate in the disk, the scaling relations from the scale invariance
model \citep{heinz:03a} are unaltered and we will not discuss them any
further, apart from reminding the reader that they provide the best
fit to the observed correlation, provided that the disk emission is
radiatively inefficient, \citep[with $L_{\rm x} \propto
{\dot{m}}^2$][]{merloni:03}.

If the X-rays originate in the jet, however, we must include radiative
cooling, based on the arguments presented in \S\ref{sec:cooling2}.
The presence of a cooling break in the spectrum was already discussed
in \S\ref{sec:cooling}.  We define the magnitude of the break as
$\Delta\alpha_{\rm x} \equiv \alpha_{\rm x} - (p-1)/2$.  We will
assume in the following that this break lies below the X-ray band for
{\em all} sources of interest.  If this were not the case, the break
would move through the X-ray band and produce an observable change in
the correlation coefficients.  Given the scatter in the ``fundamental
plane'' relation, we cannot rule out such a change with any certainty.
We note that the presence of such a change would be an indication of
jet X-rays.  The dependence of the break frequency $\nu_{\rm b}$ on
$M$, $\dot{m}$ (or alternatively, $F_{\rm x}$), and the radio
luminosity $L_{\rm r}$ can be easily calculated from
eqs.~(\ref{eq:gamma}) and (\ref{eq:nucrit}) and compared to future
data.

We will now use eqs.~(\ref{eq:xirm}) and (\ref{eq:xirx}) to test
whether radiative cooling will improve the fit of synchrotron X-rays
to the ``fundamental plane'' correlation.  In the simplest possible
case of continuous injection of fresh powerlaw particles with index
$p$ (e.g., by a stationary shock) without any adiabatic losses, the
spectral break produced by cooling is $\Delta \alpha_{\rm x} = 1/2$.
Thus, for a canonical synchrotron spectrum of index $p=2$, $\alpha =
1/2$, and $\Delta \alpha_{\rm x} = 1/2$, the steepened spectrum above
the break has a powerlaw index of $\alpha_{\rm x} = 1$. Inserting this
value into eqs.~(\ref{eq:xirm}) and (\ref{eq:xirx}) gives $F_{\rm x}
\propto M\dot{m}$ and thus $\xi_{\rm RM} = 0$ and $\xi_{\rm RX} = 1.42
$.

This implies that cooled synchrotron emission from the jet follows the
exact same scaling relation as efficient disk X-rays.  As shown in
\cite{merloni:03}, radiatively efficient disk X-rays are inconsistent
with the observed correlation coefficients of the ``fundamental
plane''.  This is clear from comparing $\xi_{\rm RM}=0$ and $\xi_{\rm
RX}=1.42$ with the measured correlation coefficients of $0.78$ and
$0.6$ respectively.  Thus, jet X-rays, in the most straightforward
scenario of synchrotron emission (which includes cooling and the
canonical range of particle spectral indices), are {\em not}
compatible with the observed correlation.

We can ask the broader question under what conditions the scaling
relations predicted by synchrotron X-rays are actually consistent with
the observed correlation coefficients.  We will do this by fitting the
``fundamental plane'' data set with the correlation coefficients
$\xi_{\rm RM}$ and $\xi_{\rm RX}$ from eqs.~(\ref{eq:xirm}) and
(\ref{eq:xirx}).  We will ignore upper limits in the data and use the
same merit chi-square estimator employed in
\cite{merloni:03,gebhardt:00}.  We also use the same error estimate,
i.e., we assume isotropic errors in all variables and normalize them
such that the reduced chisquare of the best fit is one.  This is
equivalent to equating the scatter in the fundamental plane to the
uncertainty and is a conservative approach for calculating errors.

\begin{figure}
\resizebox{\columnwidth}{!}{\includegraphics{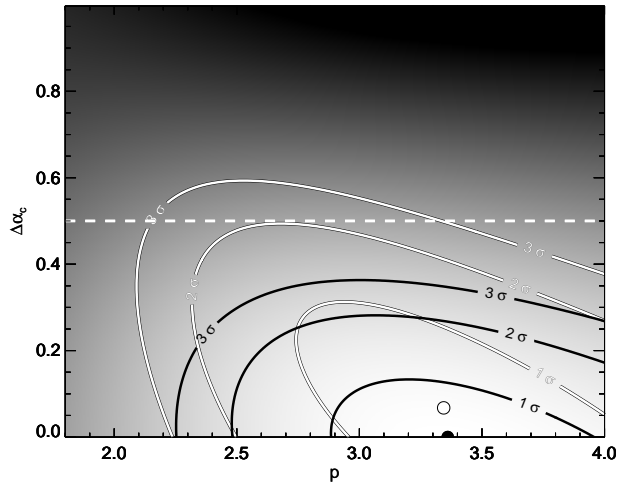}}
\caption{Chi-square map of the particle spectral index $p$ and
  synchrotron cooling break$\Delta\alpha_{\rm c}$, derived by fitting
  the ``fundamental plane'' radio-X-ray-mass correlation with cooled
  synchrotron emission as the source of the X-rays and a flat,
  optically thick radio synchrotron spectrum with index $\alpha_{\rm
  r}=0$.  Over-plotted are the the 1, 2, and 3-sigma contours (thick
  black lines), the best fit (black dot), and the canonical cooling
  break of $\Delta\alpha_{\rm c}=1/2$ (white dashed line).  A
  reasonable fit requires an unbroken, i.e. {\em uncooled} global
  spectrum that is steeper than $p\gtrsim 3$.  Also shown are the same
  1, 2, and 3 sigma contours and the best fit obtained by fitting only
  low-X-ray luminosity sources with $L_{\rm x} \leq 10^{-3}L_{\rm
  Edd}$ (white contours and white dot).
  \label{fig:alpha_p}}
\end{figure}

Figure \ref{fig:alpha_p} shows a chi-square map for the two
interesting parameters $\Delta\alpha_x$ and $p$ (marginalized over the
unknown relative normalization between radio and X-ray flux).  As
expected, the figure shows that a standard cooled synchrotron spectrum
with $p=2$ and a cooling break of $\Delta \alpha_{\rm c} = 0.5$ is
inconsistent with the correlation coefficients derived from the
``fundamental plane'' relation by more than 3 sigma.  The best fit
actually requires a {\em steep, unbroken} powerlaw with
$p=3.4^{+0.6}_{-0.5}$, $\Delta\alpha_{\rm c}=0^{+0.13}_{-0}$ and
$\alpha_{\rm x}=1.2^{+0.3}_{-0.25}$, where the lower bound on
$\Delta\alpha_{\rm c} \geq 0$ is set by the condition that radiative
cooling cannot produce positive curvature in the spectrum over a large
range in wavelengths.  Formally, the best fit would actually require a
negative value of $\Delta\alpha$.

Such a steep value of $p$ is not in line with the particle spectra
produced by Fermi I acceleration in shocks.  Furthermore, this value
of $\alpha_{\rm x}$ is significantly larger than the mean and median
of the measured X-ray powerlaw spectra in Fig.~\ref{fig:spectra}.
Most of the sources in this figure lie to the left of the grey
1-$\sigma$ confidence limit on $\alpha_{\rm x}$ derived from
Fig.~\ref{fig:alpha_p}.  The immediate conclusion from this
discrepancy is that synchrotron cooling, which should be at work in
these sources (as argued in \S\ref{sec:cooling}) does {\em not}
improve the fit to the ``fundamental plane'' (as speculated by
\citealt{merloni:03}).  To the contrary, it actually makes it worse.

\cite{merloni:03} and \cite{maccarone:03} discuss the possibility that
the radio-X-ray-mass correlation of the ``fundamental plane'' breaks
down at high accretion rates and that different emission mechanisms
are responsible for the X-rays above and below the break.  However,
repeating the analysis with a sample limited to low X-ray luminosities
($L_{\rm x} < 10^{-3}L_{\rm Edd}$) does not change the conclusions
reached above, only the confidence contours in Fig.~\ref{fig:alpha_p}
are slightly less stringent: $p = 3.3^{+0.8}_{-0.6}$,
$\Delta\alpha_{\rm x}=0.1^{0.2}_{-0.1}$, and $\alpha_{\rm
x}=1.25^{+0.35}_{-0.3}$, as shown by the white contours in
Fig.~\ref{fig:alpha_p}.

In fact, the X-ray spectra of the low luminosity sources (white dots
in Fig.~\ref{fig:spectra}) tend to be flatter than those of the high
luminosity sources (black dots).  While all the observed spectral
indices in Fig.~\ref{fig:spectra} are from AGNs, the same is known to
be the case in XRBs, which typically show harder spectra in their low
luminosity state.  Thus, most of the observed X-ray spectral indices
of low luminosity sources are significantly flatter than the range in
$\alpha_{\rm x}$ allowed by fitting the scaling relations for
synchrotron jet X-rays to the low luminosity sources in the
``fundamental plane'' sample.

\subsection{Discussion}
This discrepancy between the correlation indices of the fundamental
plane, the predicted scaling relations, and the actually observed
X-ray spectra suggests several possible scenarios (not necessarily
mutually exclusive):

\subsubsection{Steep global spectra}
\label{sec:steepspec}
Fig.~\ref{fig:alpha_p} indicates that the fundamental plane relation
is compatible with X-ray synchrotron emission {\em if} the underlying
particle distribution is inherently steep, with $p \sim 3.3$.  Such a
steep, unbroken particle distribution can be achieved in two ways:
\vspace*{6pt} 

{\noindent {\bf a)} The particle acceleration process produces
inherently steep spectra with $p\sim 3$ (unlike Fermi I acceleration),
and radiative cooling is unimportant even at X-ray energies.  However,
for radiative cooling to have no effect, even on the X-ray emitting
electrons, the magnetic fields in the inner jet would have to be {\em
very} low.}
\vspace*{6pt} 

{\noindent {\bf b)} Particle acceleration produces spectra close to $p
\sim 2$, but radiative cooling steepens the spectra to an index of $p
\sim 3$ {\em all the way down below radio frequencies} (in which case
the formalism above cannot distinguish between cooling and an
intrinsically steep spectrum).  However, the jet is optically thick at
lower frequencies. The cooling coefficient for electrons radiating
below the self-absorption turnover frequency is significantly reduced
\citep{mccray:69,ghisellini:88}, and an unbroken powerlaw is not a
solution to the kinetic equation in this case.  The self-absorption
barrier can be circumvented if radiation losses are dominated by
inverse Compton scattering.  A radiative signature of this component
must be visible, since it would dominate the radiative output of the
jet (presumably in the X-rays).  Alternatively, a somewhat fine--tuned
adiabatic loss scheme could keep the synchrotron cooling break always
below the local turnover frequency.}
\vspace*{6pt}

However, such a steep global spectrum is inconsistent with the flatter
radio spectral indices observed in larger scale optically thin jet
sources (like M87, GRS 1915+105, 3C273, etc.) and would therefore
require additional particle acceleration on larger spatial scales.  It
is also inconsistent with the standard scenario for producing flat
radio spectra \citep{blandford:78}.  And for most sources in
Fig.~\ref{fig:spectra} (and XRBs in the low-hard state), the observed
X-ray spectral indices $\alpha_{\rm x}$ are much shallower than this.
Ultimately, broad band spectra of the jet core are necessary to verify
the spectral index of the optically thin part of the spectrum over a
large frequency range and for a number of sources with different $M$
and $\dot{m}$ to determine $p$ unequivocally.

\subsubsection{Scale invariance violated}
\label{sec:notscaleinvariant}
It is possible that the X-rays of all sources in the sample are of
synchrotron--jet origin, but that jets are inherently not scale
invariant and the analysis presented above and in \citet{heinz:03a}
does not apply.  However, the presence of a strong correlation and its
agreement with the scale invariance predictions would be a
coincidence.  Moreover, specific models used by \citet{markoff:01} and
\citet{markoff:03} \citep[based on the original, scale invariant jet
model by][]{blandford:79} and by \citet{falcke:95},
\citet{markoff:03}, and \citet{falcke:04} to model the correlation and
the broad band spectra of individual sources are themselves members of
the class of scale invariant models. Such a lack of scale invariance
would require a more complex approach at understanding jet spectra and
correlations.

\subsubsection{Other sources of X-ray emission}
\label{sec:diskxrays}
Finally, the discrepancy can be resolved if the X-ray emission from at
least a significant fraction of the sources in the ``fundamental
plane'' sample does not originate as synchrotron emission in the jet,
but either a) in the disk, as long as the X-ray emission from the disk
is inefficient in the sense that $F_{\rm x} \propto \dot{m}^2$, or b)
as inverse Compton emission from the jet, with seed photons provided
by the accretion flow, the jet radio synchrotron emission, or the
companion \citep{georganopoulos:02,markoff:04}.

The disk interpretation would be in line with the fact that the
inefficient disk X-ray scenario provides the best global fit to the
fundamental plane relation.  Discussing the inverse Compton scenario
is beyond the scope of this paper.  It is clear, though, that the
presence of the radio-X-ray-mass correlation will impose similar
limits on a Compton scenario as is the case in disk X-rays (e.g., it
might help us to determine the source of the up-scattered photons,
i.e., disk vs. synchrotron-self-Compton).  Clearly, further research
is necessary to explore this scenario.

Given the large scatter in the fundamental plane relation, the X-rays
of at least a part of the sample might still be predominantly
jet--synchrotron emission.  In fact, it is quite plausible to expect
that all three emission mechanisms will make a contribution to the
X-ray spectrum of any given source at some level.  It might well be
that in different sources, different X-ray emission mechanisms
dominate, leading to the significant observed scatter in the
``fundamental plane''.  For example, the X-rays from Galactic XRBs in
the low hard state might be of jet synchrotron origin, while the
X-rays from AGNs might originate primarily in the disk (or vice
versa). In this case, the fact that both XRBs and AGNs fall on {\em
the same} relation would be a coincidence\footnote{The fact that both
populations follow very similar correlations between $L_{\rm r}$ and
$L_{\rm x}$ for a fixed mass bin would be a consequence of the fact
that the predicted scaling relations from \citet{heinz:03a} are
relatively similar for inefficient disk X-rays and uncooled
synchrotron X-rays from the jet.}.

The scaling relations derived above and in \cite{heinz:03a,falcke:95}
can help to distinguish between these different types of sources in a
large enough, well characterized sample, which is clearly a primary
objective for future research in this area.

\subsection{Caveats}
Some caveats and sources of uncertainty should be kept in mind when
interpreting these results:
\vspace*{6pt}

{\noindent {\bf a)} The ``fundamental plane'' sample is a relatively
diverse sample, containing several different types of AGNs, spanning a
large range in accretion rate and black hole mass (see
\citealt{merloni:03} for a detailed discussion of the properties of
the sample), mostly for the sake of a large number of sources and for
testing the universality of the correlation.  This is in line with the
suggestion from \S\ref{sec:diskxrays} that sources with different
X-ray emission mechanisms may be present in the sample.  Due to the
diverse nature of the sample, some systematic effects may be present
that we have not accounted for.  In the context of this paper,
however, a reduction to the more homogeneous low-luminosity sample
(also used by \citealt{falcke:04}) shows identical results.}
\vspace*{6pt}

{\noindent {\bf b)} The X-ray spectral characteristics of XRBs change
as they change state, and the radio and X-ray band vary on different
time scales and with respective phase lags. This specifically affects
the global spectral index.  The same should hold for AGNs, though it
is difficult to test due to the long time scales involved.  However,
as long as there is a simple delay between the two bands and no
systematic trend for one band (X-ray or radio) to be in a specific
state longer than the other, the variability should average out in a
large enough sample and be simply another source of scatter in the
fundamental plane. The presence of the radio-X-ray correlation in XRBs
in GX 339-4 \citep{corbel:03} and V404 Cyg \citep{gallo:03} indicates
that variability effects do not destroy the correlation at least in
the case of XRBs.  It has to be kept in mind that the error bars on
some of the $\alpha_{\rm x}$ measurements from the literature shown in
Fig.~\ref{fig:spectra} are very large.  Clearly, future observations
with tighter error bars will be very helpful to strengthen the
arguments made in this paper.}
\vspace*{6pt}

{\noindent {\bf c)} The same holds for the radio spectral index: it is
known that as a the source flux changes in XRBs, $\alpha_{\rm r}$ also
varies by significant amounts.  However, at least for AGNs we know
that, on average, a flat spectrum is a good description of the radio
core, and while the situation for XRBs at very low luminosities should
be explored further, it is probably a good approximation to assume
that the spectra are, on average, close to flat \citep{fender:01c}.
Again, the radio-X-ray correlation indicates that the underlying
spectrum is unlikely to vary randomly, since the radio-X-ray
correlation would otherwise be a coincidence.}
\vspace*{6pt}

{\noindent Thus, while we expect all of these effects to be present
and to be important on a detailed, quantitative level, we believe that
they do not alter the fundamental presence of the correlation and that
we can use a statistical approach (i.e., measuring the correlation
coefficient) to learn something about the underlying nature of the
emitter.  However, all of these effects will contribute to the scatter
and the uncertainty in any conclusion and must therefore be better
characterized in future studies.}

\section{Conclusions}
\label{sec:conclusions}
We presented an extension of the scale invariance formalism derived by
\citet{heinz:03a} that takes the effects of radiative cooling into
account self-consistently.  The spectral index information of the
sources contributing to the ``fundamental plane'' correlation shows
that the X-ray luminosity of the sources should be affected by cooling
if it is due to synchrotron emission from the jet.  We proved that,
for the canonical synchrotron parameters ($p=2$, $\Delta\alpha =
1/2$), the scaling relations for X-ray and radio synchrotron emission
cannot reproduce the correlation coefficients found by
\citet{merloni:03}.

Fitting the newly derived scaling relations to the fundamental plane
data, we showed that steep ($p \sim 3.4$, $\alpha_{\rm x} \sim 1.2$),
{\em unbroken} spectra are required to reproduce the correlation
coefficients.  Such steep spectra are inconsistent with the observed
mean and median spectral index of the ``fundamental plane'' sources
with measured spectra.  This seeming discrepancy leads us to propose
three alternative scenarios: a) The X-rays are of synchrotron origin
and the entire electron spectrum has a steep index of order $\sim 3$,
which would require steep injection spectra or inverse Compton cooling
of the entire powerlaw distribution of electrons to well below radio
frequencies.  However, such steep spectra are inconsistent with the
observed range in X-ray spectral indices $\alpha_{\rm x}$; b) the
X-rays are due to synchrotron emission from jets, but some of the
assumptions made in deriving the scale invariance predictions fail
(i.e., some aspect of jet dynamics is not scale invariant); c) the
X-rays from a sizeable fraction of the sources in the sample are not
of jet-synchrotron origin and are thus most likely emitted either by
the accretion flow itself or as Compton up-scattered radiation from
the jet.  It still plausible that a fraction of the sources (e.g.,
most of the XRBs) in the sample emit predominantly synchrotron X-rays
and that synchrotron emission from jets contributes a small fraction
to the X-ray spectrum of most sources.

The analysis presented in this paper shows how a statistical analysis
of radio and X-ray luminosities of a sample of black holes can be used
to constrain the X-ray emission mechanism of accreting black holes and
to test the scale invariance hypothesis of jets.  A more refined
sample of black holes with a better understanding of the sources of
scatter (and the selection effects involved), and with tighter limits
on the X-ray spectral indices is necessary to take this analysis to a
more precise quantitative level.
\vspace*{24pt}

\thanks{We would like to thank Sera Markoff, Andrea Merloni, Mike
  Nowak, and Rashid Sunyaev for helpful discussions.  Support for this
  work was provided by the National Aeronautics and Space
  Administration through Chandra Postdoctoral Fellowship Award Number
  PF3-40026 issued by the Chandra X-ray Observatory Center, which is
  operated by the Smithsonian Astrophysical Observatory for and on
  behalf of the National Aeronautics Space Administration under
  contract NAS8-39073.}

\appendix
\section{Full derivation of the scaling relations}
\subsection{Particle transport}
\label{sec:appendix}
In this appendix, we will derive the scaling relations for X-ray
synchrotron radiation from scale invariant jets, including radiative
cooling.  We refer the reader to \S\ref{sec:cooling} for the set of
assumptions going into the scale invariance ansatz.  Again, we assume
that the particles are injected into the jet at some scale invariant
location $r_0$ and time $t_0$ (the latter measured in the jet frame).

The kinetic equation describing the evolution of the particle
distribution function $f(\gamma)$ as a function of time $t$ and
particle energy $\gamma$ is \citep[adapted from][]{coleman:88}:
\begin{equation}
  \frac{\partial f}{\partial t} +
  \frac{d\ln{(\rho)}}{dt}\left[\frac{\gamma^3}{3}\frac{\partial}{\partial
  \gamma}\left(\gamma^{-2}f\right)\right] = \frac{2 e^4 \langle
  B_{\perp}^2 \rangle}{3 m_{\rm e}^3 c^5}\frac{\partial{\left(\gamma^2
  f\right)}}{\partial \gamma}
  \label{eq:transport}
\end{equation}
where an isotropic pitch angle distribution of the particles was
assumed for the numerical coefficients (this assumption is not
critical for the derivation, since the relevant numerical coefficients
don't enter into the scaling relations).  Eq.~(\ref{eq:transport}) can
be easily expanded to include inverse Compton losses by adding a term
proportional to the radiation energy density to the right hand side of
eq.~(\ref{eq:transport}).

For mathematical convenience, we substitute 
\begin{equation}
  \hat{\gamma} \equiv \left(\rho_0/\rho\right)^{1/3}\gamma
\end{equation}
where $\rho_0 \equiv \rho(t_0)$ is the density at the point
$\chi_0=r_0/r_{\rm g}$ where the particles were accelerated/injected,
and $\tilde t\equiv t/t_0$ is the time when the particles are injected
(measured in the jet frame).  We also substitute $\tilde{f} \equiv
\left(\rho/\rho_0\right)^{-2/3} f$.

As plasma travels along the jet, there is a one-to-one relation
between its jet coordinate $\chi$ and the time $\tilde t$, defined by
\begin{equation}
  \frac{d\chi}{d{\tilde t}} = \frac{t_0}{r_{\rm g}}{c\beta\Gamma} =
  \frac{t_0 c}{r_{\rm g}}{\psi_{\beta\Gamma}(\chi)}
\end{equation}
We define $\chi_0$ as the acceleration location such that ${\tilde
t}(\chi_0)=1$.  By virtue of the assumption that the particle
injection location is scale invariant, this value of $\chi_0$ must be
independent of $M$ and $\dot{m}$.  We can then substitute the
integration over $\tilde t$ by one over $\chi$.

The characteristic equation for synchrotron and adiabatic losses for
each particle implicit in eq.~(\ref{eq:transport}) is:
\begin{subequations}
\begin{eqnarray}
  \hspace{-12pt}\lefteqn{\frac{d\hat{\gamma}}{d\chi} = -{A_0}
  \frac{1}{\psi_{\beta\Gamma}(\chi)}
  \left(\frac{B}{B_0}\right)^2\left(\frac{\rho}{\rho_0}\right)^{1/3}
  {\hat{\gamma}}^2} \ \ \ \ \
  \label{eq:characteristic}
  \\ & = & - {A_0} \left[\psi_{\beta\Gamma}(\chi)\right]^{-1}
  \left[\psi_{B}(\chi)\right]^2
  \left[\psi_{\rho}(\chi)\right]^{1/3}{\hat{\gamma}}^2 \\ A_0 &
  \equiv & r_g {\phi_{B}}^2 \frac{4 e^4}{9 m^3 c^6}
\end{eqnarray}
\end{subequations}
where we used eq.~(\ref{eq:scaleinvariant}) in the last step.

Equation (\ref{eq:characteristic}) has the solution
\begin{eqnarray}
  \hat{\gamma}(\chi) = \frac{1}{\frac{1}{{\hat{\gamma}}_0} + A_0
  \int_{\chi_0}^{\chi} d{\chi}' \frac{\left[\psi_B(\chi')\right]^2
  \left[\psi_{\rho}(\chi')\right]^{\frac{1}{3}}}
  {\left[\psi_{\beta\Gamma}(\chi')\right]}}
  \label{eq:gamma}
\end{eqnarray}
where ${\hat{\gamma}}_0 \equiv \gamma(t_0)$.  For convenience we
define
\begin{subequations}
\begin{eqnarray}
  {\xi(\chi)} & = & \frac{A_0 \int_{\chi_0}^{\chi} d\chi'
\left[\psi_{\beta\Gamma}(\chi')\right]^{-1}
\left[\psi_B(\chi')\right]^2 \left[\psi_{\rho}(\chi')\right]^{1/3}}{
\left[\psi_{\rho}(\chi)\right]^{\frac{1}{3}}}
  \label{eq:xi} \\
  & = & \phi_{\xi}(M,\dot{m}) \psi_{\xi}(\chi)
\end{eqnarray}
\end{subequations}
where, by definition,
\begin{equation}
  \phi_{\xi}(M,\dot{m}) = M \left[{\phi_{B}(M,\dot{m})}\right]^2
  \label{eq:phi_xi}
\end{equation}
and $\psi_{\xi}$ is independent of $M$ and $\dot{m}$ (but might have
an implicit dependence on parameters such as black hole spin).  $\xi$
is simply the inverse of the synchrotron cutoff energy,
$\xi=1/\gamma_{\rm c}$, while $\phi_{\xi}$ describes how this energy
cutoff in the electron spectrum depends on $M$ and $\dot{m}$.

We can then solve eq.~(\ref{eq:transport}) along the characteristics
defined by eq.~(\ref{eq:gamma}), which reduces the problem to an ODE:
\begin{equation}
  \frac{d\left\{\left[{\hat{\gamma}}(\chi)\right]^2\tilde{f}
  \left[\hat{\gamma}(\chi)\right]\right\}} {d\chi} = 0
\end{equation}
where $d/d\chi$ is the total derivative along curves
$\hat{\gamma}(\chi)$ defined by eq.~(\ref{eq:gamma}).  In other words,
the function ${\hat{\gamma}}^2\tilde{f}$ is constant along solutions
of eq.~(\ref{eq:characteristic}).  The well known solution to
eq.~(\ref{eq:transport}) is then \citep{kardashev:62,bicknell:82}:
\begin{equation}
  f(\gamma,\chi) =
  \left(\frac{\left[\psi_{\rho}(\chi)\right]^{-\frac{1}{3}}{\gamma}}{1
  - \xi\gamma},\chi_0\right)
  \left(1 -
  \xi{\gamma}\right)^{-2}\left[\psi_{\rho}(\chi)\right]^{\frac{2}{3}}
\end{equation}
for $\gamma \leq \gamma_{\rm c} \equiv 1/\xi$ and the solution has a
cutoff at $\gamma_{\rm c}$ such that $f(\gamma > \gamma_{\rm c},t) =
0$.  For later use, we define the scaled Lorentz factor as the ratio
of the particle energy to this cutoff:
\begin{equation}
  \tilde \gamma \equiv \gamma \xi = \frac{\gamma}{\gamma_{\rm c}}
\end{equation}
Using the assumption that the injected particle distribution at
$\chi_0$ is a powerlaw as described in eq.~(\ref{eq:powerlaw}), we can
write $f(\gamma,\chi)$ as
\begin{equation}
  f(\gamma,\chi) = \phi_{C} \xi^{p} {\hat \gamma}^{-p} \left( 1 - \hat
  \gamma\right)^{p - 2} \left[\psi_{\rho}(\chi)\right]^{(p + 2)/3}
  \label{eq:distributionsolution}
\end{equation}

\subsection{Synchrotron emission from cooled particle distributions}
Following the notation in \citet{rybicki:79}, we define the critical
synchrotron frequency as
\begin{eqnarray}
  \nu_{\rm c}(\gamma) & \equiv & \frac{3e}{16mc} B \gamma^2
  \label{eq:nucrit}
\end{eqnarray}
(where we have averaged over the assumed isotropic pitch angle
distribution -- as before, this assumption is not critical for the
derivation below).  This is the frequency where most of the energy of
the synchrotron kernel is emitted for an electron of energy $\gamma
m_e c^2$.

The synchrotron luminosity $F_{\nu}$ at frequency $\nu$ is then
\begin{eqnarray}
  \lefteqn{F_{\nu}=} \hspace{0.92\columnwidth} \\ \ \ \ \ \
  \int_{r(t_0)}^{\infty}dr \pi R_{\rm jet}^2 \frac{\sqrt{3} e^3 B
  \langle sin{ \alpha } \rangle}{m_{\rm e}c^2} \int_{\gamma_{\rm
  min}}^{\gamma_{\rm c}}d\gamma f(\gamma) {\rm
  F_{s}}\left(\frac{\nu}{\nu_c(\gamma)}\right) \nonumber
\end{eqnarray}
where
\begin{equation}
  {\rm F_{s}}(x) \equiv x \int_{x}^{\infty} dx' {\rm K_{5/3}}(x')
\end{equation}
is the synchrotron kernel.  Substituting the expressions from
eq.~(\ref{eq:distributionsolution}) and eq.~(\ref{eq:scaleinvariant})
we get
\begin{eqnarray}
  \lefteqn{F_{\nu} = M^3 \phi_{B} \phi_{C} {\phi_{\xi}}^{p-1} F_0
  \int_{\chi_0}^{\infty}d\chi \left[
  \psi_{B}\psi_{R}^2\psi_{\rho}^{(p+2)/3}\psi_{\xi}^{p-1}
  \raisebox{12pt}[12pt][12pt]{\ }\right.}  \label{eq:fnu} \\ & &
  \hspace{0.30\columnwidth} \left. \raisebox{12pt}[12pt][12pt]{\ }
  \int_{0}^{1}d{\tilde \gamma}\ {\tilde \gamma}^{-p}\left(1 - \tilde
  \gamma\right)^{p-2}{\rm F}\left(\frac{\tilde \nu
  {\psi_{\xi}}^2}{\psi_{B}{\tilde\gamma}^2}\right)\right] \nonumber
\end{eqnarray}
where we have defined
\begin{equation}
  \tilde \nu \equiv \frac{16m_{\rm e}c}{3e} \frac{\nu\,
  {\phi_{\xi}}^2}{\phi_{B}}
\end{equation}
and 
\begin{equation}
  F_0 \equiv \frac{\pi^2 \sqrt{3} e^3}{4 m_{\rm e} c^2}
\end{equation}
(where we have taken the magnetic field to be isotropically tangled,
without loss of generality).

Before progressing further, it is important to note several things
about eq.~(\ref{eq:fnu})
\begin{itemize}
\item{The low energy cutoff enters into the boundaries of the integral
  over $\tilde \gamma$. However, since we are dealing with optically
  thin synchrotron radiation at high frequencies (X-rays), the low
  energy cutoff has a negligible effect on the integral, since the
  synchrotron kernel drops exponentially at low particle energies and
  the integral converges even if the lower bound is placed at zero,
  which we have done in eq.~(\ref{eq:fnu}).}
\item{As a function of $\tilde \nu$, the integrand is entirely
  independent of $M$ and $\dot{m}$, which enter only into $\phi_{B}$,
  $\phi_{C}$, and $\phi_{\xi}$ in front of the integral. However,
  $\tilde \nu$ {\em does} depend on $M$ and $\dot{m}$ through
  $\phi_{B}$ and $\phi_{C}$.}
\item{As discussed earlier, $\phi_{B}$ and $\phi_{C}$ are simply the
  magnetic field strength $B$ and the normalization $C$ of the
  particle distribution at the base of the jet (both of which vary
  with $M$ and $\dot{m}$), whereas $\phi_{\xi}$ describes how the
  cutoff energy $\gamma_{\rm c}$ at a given position $\chi$ depends on
  $M$ and $\dot{m}$, as defined in eq.~(\ref{eq:phi_xi}).  The fact
  that we can write down a simple dependence of $\gamma_{\rm c}$ on
  $M$ and $\dot{m}$ will allow us below to set the cooled
  synchrotron flux in relation to $M$ and $\dot{m}$ as well.}
\end{itemize}

Following the same procedure as \citet{heinz:03a}, we are now in a
position to derive the scaling relation between the synchrotron flux
at X-ray energies $F_{\nu}$ and $M$:
\begin{eqnarray}
  \lefteqn{\frac{\partial \log{F_{\nu}}}{\partial \log{M}} =
  \frac{\partial
  \log{\left(M^3\phi_{B}\phi_{C}\phi_{\xi}^{p-1}\right)}}{\partial
  \log{M}} + \frac{\partial \log{F_{\nu}}}{\partial \log{\tilde
  \nu}}\frac{\partial \log{\tilde \nu}}{\partial \log{M}}} \nonumber
  \\ & = & 2 - 2\alpha_{\rm x} + p +\frac{\partial
  \log{(\phi_{C}})}{\partial \log{M}} + (2p - 1 - 3\alpha_{\rm
  x})\frac{\partial \log{(\phi_{B})}}{\partial \log{M}} \nonumber \\
  & \equiv & \xi_{\rm xM}
  \label{eq:axim}
\end{eqnarray}
where we used the X-ray spectral index $\alpha_{\rm x}$ such that
$F_{\rm x} \propto \nu^{-\alpha_{\rm x}}$ or $\alpha_{\rm x} \equiv
-\frac{\partial \log{F_{\nu}}}{\partial \log{\nu}}$.

Note that eq.~(\ref{eq:xim}) is fully general as long as synchrotron
self-absorption is negligible, even in the absence of any cooling.
This can be checked by comparing this eq.~(\ref{eq:xim}) to the
equivalent expression in \citet{heinz:03a} by substituting
$\alpha_{\rm x} = (p-1)/2$, which is the appropriate expression of
optically thin, uncooled synchrotron emission.

Similarly, we can derive the scaling relation between $F_{\nu}$ and
the accretion rate $\dot{m}$:
\begin{eqnarray}
  \lefteqn{\frac{\partial \log{F_{\nu}}}{\partial \log{\dot{m}}} =
  \frac{\partial
  \log{\left(M^3\phi_{B}\phi_{C}\phi_{\xi}^{p-1}\right)}}{\partial
  \log{\dot{m}}} + \frac{\partial \log{F_{\nu}}}{\partial \log{\tilde
  \nu}}\frac{\partial \log{\tilde \nu}}{\partial \log{\dot{m}}}}
  \nonumber \\ & = & \frac{\partial \log{(\phi_{C}})}{\partial
  \log{\dot{m}}} + (2p - 1 - 3\alpha_{\rm x})\frac{\partial
  \log{(\phi_{B})}}{\partial \log{\dot{m}}} \equiv \xi_{\rm x\dot{m}}
  \label{eq:aximdot}   
\end{eqnarray}
$\dot{m}$ is simply a parameter describing the variation of magnetic
field and particle pressure at the base of the jet at constant $M$,
and must not necessarily be directly proportional to the accretion
rate. These two equations are the scaling laws used throughout this
paper

The superposition of cooled electron spectra from different regions in
the jet might, in principle, produce any number of spectral shapes
(e.g., broken powerlaws, cutoff, curved spectra, etc.).
Eqs.~(\ref{eq:xim}) and (\ref{eq:ximdot}), as derived in the appendix,
are independent of what the actual X-ray spectral shape is\footnote{In
situations where the spectrum is complicated, $\alpha_{\rm x}(\nu)$
will itself depend on $M$ and $\dot{m}$ and eqs.~(\ref{eq:xim}) and
(\ref{eq:ximdot}) are then differential equations between $F_{\nu}$
and $M$ and between $F_{\nu}$ and $\dot{m}$.  Solving them requires
knowledge of the synchrotron spectrum over a large enough frequency
range to cover the desired range in $M$ and $\dot{m}$.}.  However,
observations of synchrotron spectra from sources where we know cooling
is efficient show either exponential-type cutoffs or, if fresh plasma
is continuously supplied, broken powerlaw spectra. The latter case is
appropriate here, since we are observing a superposition of spectra
from plasma that moves away from a stationary or quasi-stationary
acceleration region.  Therefore, we can expect the cooled X-ray
spectrum to be of powerlaw type:
\begin{equation}
  F_{\rm x}\propto M^{\xi_{rm xM}}{\dot{m}}^{\xi_{\rm x\dot{m}}}
\end{equation}

\subsection{Inverse Compton cooling}
It is easy to change the cooling mechanism in eq.~(\ref{eq:transport})
to inverse Compton cooling off of radiation from either the central
accretion disk or the microwave background.  The latter will be
entirely irrelevant in the context of core dominated jets and we will
ignore it in the following paragraph.  The former can be included by
substituting $8\pi\,U_{\rm rad}$ for ${B_{\perp}}^2$ in
eq.~(\ref{eq:transport}). In this case the function $\phi_{\xi}$ will
have to be changed to reflect the dependence of the radiation energy
density on $M$ and $\dot{m}$, which depends on the accretion
scenario. In the case of an efficient accretion disk with $\phi_{\rm
B}=\sqrt{\dot{m}/M}$, we have
\begin{equation}
  \phi_{\xi} \propto r_{\rm g}M^{-1}\dot{m}
\end{equation}
which has exactly the same dependence on $M$ and $\dot{m}$ as in the
synchrotron loss case, and the solution is unchanged.  In radiatively
inefficient flows, we have approximately
\begin{equation}
  \phi_{\xi} \propto r_{\rm g}M^{-1}{\dot{m}}^2
\end{equation}
which implies that inverse Compton is increasingly unimportant
compared to synchrotron cooling at low accretion rates.  It also
implies that, while $\xi_{\rm xM}$ is unchanged, $\xi_{{\rm
x}\dot{m}}$ is changed to
\begin{equation}
  \xi_{{\rm x}\dot{m}} = 2p - 1/2 -3\alpha_{\rm x}
\end{equation}

\end{document}